\documentclass[preprint]{aastex631}
\usepackage{natbib}
\received{July 6, 2022}
\accepted{August 10, 2022}

\submitjournal{ApJ}

\begin{document}

\newcommand{\ltsimeq}{\raisebox{-0.6ex}{$\,\stackrel{\raisebox{-.2ex}%
{$\textstyle<$}}{\sim}\,$}}
%
\newcommand{\gtsimeq}{\raisebox{-0.6ex}{$\,\stackrel{\raisebox{-.2ex}%
{$\textstyle>$}}{\sim}\,$}}

                                                                                                                                        
\title{HST UV Spectroscopy of the Planet-Hosting T Tauri Star PDS 70}

\correspondingauthor{Stephen L. Skinner}
\email{stephen.skinner@colorado.edu}

\author{Stephen L. Skinner}
\affiliation{Center for Astrophysics and
Space Astronomy (CASA), University of Colorado,
Boulder, CO, USA 80309-0389}

\author{Marc Audard}
\affiliation{Department of Astronomy, University of Geneva,
Ch. d'Ecogia 16, 1290 Versoix, Switzerland}


\begin{abstract}
\small{We summarize {\em Hubble Space Telescope} ({\em HST}) UV observations of the
weak-lined T Tauri star (wTTS) PDS 70 obtained with the Space Telescope Imaging
Spectrograph (STIS). These observations provide the first far-UV (FUV) and
near-UV (NUV) spectra of PDS 70. Ground-based observations have so far
revealed two formative giant planets orbiting in a wide gap in its circumstellar
disk. Both the star and young planets are thought to still be accreting.
The {\em HST} spectra provide new insight into physical conditions in the
star's outer atmosphere and  circumstellar environment. The spectra
are dominated by chromospheric and transition region emission lines with maximum
formation temperatures log T = 4.5 - 5.2 K.  Stellar continuum emission is
present in the NUV but we find no significant FUV continuum, as could arise from
accretion shocks. Several fluorescent FUV H$_{2}$ emission lines are present,
a surprising result since H$_{2}$ lines are usually undetected in wTTS. The
H$_{2}$ lines likely originate in irradiated circumstellar gas that
could serve as a reservoir for the star's waning accretion.
A previously established correlation between C IV line luminosity and accretion 
rate yields $\dot{M}_{acc}$ $\sim$  10$^{-10}$
$M_{\odot}$ yr$^{-1}$, consistent with previous estimates.
ALMA disk gas models imply strong absorption of stellar
X-ray and UV (XUV) radiation near the star, effectively shielding
the planets. Inner disk gas is exposed to ongoing
photoevaporation by XUV radiation and the disk is
nearing the end of its expected lifetime, making PDS 70 an important example
of a young planet-hosting star in the late stage of accretion.}
\end{abstract}
\section{Introduction}
In a remarkable discovery, Keppler et al. (2018; 2019)
reported the {\em direct detection} of a giant protoplanet named
PDS 70b orbiting the $\sim$5 Myr old TTS PDS 70 (Table 1).
It was detected at a projected angular separation of $\approx$0.$''$195
($\approx$22 au) from the star and lies within a wide dust-depleted gap 
in the circumstellar disk. The existence of PDS 70b was confirmed 
by Haffert et al. (2019) who also reported the detection of a second 
giant protoplanet PDS 70c at a projected angular separation of 
$\approx$0.$''$24 ($\approx$27 au). Assuming that the planets move in
circular orbits in the circumstellar disk and a disk inclination 
$i_{disk}$ = 49.7$^{\circ}$, Haffert et al. computed deprojected
orbital radii of $\approx$21-22 au (PDS 70b) and $\approx$34-40 au (PDS 70c). 
Their estimated masses are 4-17 M$_{Jup}$ (PDS 70b) and 4-12 M$_{Jup}$ (PDS 70c).
Both protoplanets were detected at H$\alpha$  and sub-mm wavelengths 
and are thought to still be accreting from circumplanetary disks
at low rates of (1 - 2) $\times$ 10$^{-8}$ M$_{Jup}$ yr$^{-1}$
(Christaens et al. 2019; Haffert et al. 2019; Isella et al. 2019; 
Zhou et al 2021). Active accretion implies the planets are
still in the formative stage.

Previous studies have reached different conclusions as to whether the 
star itself is still accreting. PDS 70 was  classified as a
weak-lined T Tauri star (wTTS) on the basis of its 
H$\alpha$ equivalent width (EW) by Gregorio-Hetem \& Hetem (2002).
Generally, accretion from circumstellar disks is believed to have 
ceased in wTTS (or dropped below detectable levels) 
but H$\alpha$ EW alone does not provide a reliable accretion indicator 
(Johns-Krull et al. 2000).  Using {\em Swift} UV filter
fluxes, Joyce et al. (2020) concluded that the UV emission of PDS 70 is mainly
chromospheric and derived an accretion rate no larger than
log $\dot{\rm M}_{acc}$ = $-$11.22 M$_{\odot}$ yr $^{-1}$.
Using ALMA observations obtained in 2016 with a synthesized
beam of $\approx$0$''$.17, Long et al. (2018) found no significant
gas inside of 1 au, suggesting that the star is not actively accreting.
But Haffert et al. (2019) noted that the {\em stellar} H$\alpha$ line
has an inverse P-Cygni profile showing redshifted absorption, which usually
signifies accretion in TTS. Furthermore, disk models analyzed by Manara et al. (2019)
found PDS 70 to be accreting at a rate log $\dot{\rm M}_{acc}$ = $-$10.26
M$_{\odot}$ yr $^{-1}$.
More recently, Thanathibodee et al. (2020) modeled the H$\alpha$ line
profile of PDS 70 and concluded that the star is still accreting at
a variable rate log $\dot{\rm M}_{acc}$ = ($-$10.22 - $-$9.66) M$_{\odot}$ yr $^{-1}$.
They also detected blueshifted absorption in the He I line (10830 \AA)
at a velocity $v$ $\approx$ $-$85 km s$^{-1}$ indicating a wind
which may originate in the inner disk.
Considering the above results together, it
seems likely that the star is still accreting but at lower rates than typically
inferred for accreting classical TTS (cTTS) of similar K7 spectral type (Ingleby et al. 2013).

UV spectra provide an excellent diagnostic for characterizing physical conditions in
the  star's chromosphere and transition region, identifying accretion signatures such
as NUV or FUV excess continuum emission, and searching
for fluorescent H$_{2}$ emission from molecular gas near the star that could serve as an 
accretion reservoir. No UV spectra of PDS 70 have 
been obtained in previous work but some information on UV fluxes and X-ray emission was 
provided by {\em Swift} (Joyce et al. 2020). {\em Swift} obtained
NUV fluxes in two broad-band filters, namely the $uvw1$ filter
(2253 - 2946 \AA) and $uvw2$ filter (1599 - 2256 \AA). These
filters capture  continuum and line emission, both of which are expected to
be present in a K7-type TTS. For example, the $uvw1$ filter includes
strong lines such as the Mg II doublet (2297/2804 \AA). 
In order to distinguish between the separate line and continuum contributions 
to the total flux, UV spectra are required. In addition, {\em Swift} filters 
do not cover the FUV range below 1600 \AA~ which 
includes important diagnostic lines such as the C IV resonance doublet
(1548/1551 \AA) and many H$_{2}$ transitions.

The proximity and low visual extinction of
PDS 70 make it a superb candidate for UV spectroscopy since large
uncertainties associated with dereddening are mitigated.
Because of its importance as a young planet-hosting star that  may
still be accreting at low rates, we obtained high-quality {\em HST} 
near-simultaneous FUV and NUV grating spectra of PDS 70 in December 2020. 
These spectra characterize the stellar UV continuum and line emission of
PDS 70 for the first time and reveal fluorescent H$_{2}$ emission lines
originating in circumstellar molecular gas.

\clearpage
\begin{deluxetable}{cccccccccc}
\tabletypesize{\scriptsize}
\tablecaption{PDS 70 Stellar Properties}
\tablehead{
           \colhead{Sp. type}               &
           \colhead{Age}                    &
           \colhead{M$_{*}$}                &
           \colhead{R$_{*}$}                &
           \colhead{L$_{*}$}                &
           \colhead{T$_{\rm eff}$}              &
           \colhead{V}                      &
           \colhead{A$_{V}$}                &
           \colhead{RV}                     &
           \colhead{d}               \\
           \colhead{}                       &
           \colhead{(My)}                   &
           \colhead{(M$_{\odot}$)}          &
           \colhead{(R$_{\odot}$)}          &
           \colhead{(L$_{\odot}$)}          &
           \colhead{(K)}                    &
           \colhead{(mag)}                      &
           \colhead{(mag)}                &
           \colhead{(km/s)}               &
           \colhead{(pc)}               
}
\startdata
 K7IVe (1,2)  & 5.4$\pm$1.0 (3) &  0.76 (3)  & 1.26 (1,3)  & 0.35 (1,5) & 3972$\pm$37 (1,2) & 12.2 (4)  & 0.05$^{+0.05}_{-0.03}$ (3) & 3.13$\pm$1.4 (6)  & 112.39$\pm$0.24 (7)    \\
\enddata
\tablecomments{References: (1) Pecaut \& Mamajek 2016
                           (2) Gregorio-Hetem \& Hetem (2002) give spectral type K5 and T$_{\rm eff}$ = 4406 K. \\  
                           (3) M\"{u}ller et al. 2018 
                           (4) Kiraga 2012
                           (5) Keppler et al. 2018 
                           (6) Radial velocity (RV): {\em Gaia} DR2;
                               a value RV = 6.0$\pm$1.5 km/s was obtained by Thanathibodee et al. (2020).
                           (7) Distance: {\em Gaia} EDR3}
\end{deluxetable}

\begin{deluxetable}{lccccccc}
\tabletypesize{\scriptsize}
\tablecaption{{\em HST} STIS/MAMA First-Order UV Grating Observatations of PDS 70}
\tablehead{
           \colhead{Grating}               &
           \colhead{Start}               &
           \colhead{$\lambda$ range\tablenotemark{\scriptsize{a}}}      &
           \colhead{Resolving}            &
           \colhead{Dispersion $\Delta\lambda$}            &
           \colhead{Scale}     &
           \colhead{Exposure}            &
           \colhead{Total Time}             \\
           \colhead{}                &
           \colhead{(UT)}         &
           \colhead{(\AA)}           &
           \colhead{Power\tablenotemark{\scriptsize{b}}}           &
           \colhead{(\AA/pixel)}     &
           \colhead{($''$/pixel)}         &
           \colhead{(s)} &
           \colhead{(s)}              
}
\startdata
 G230L (NUV) & 2020-12-29 11:09  & 1570-3180 (2376)  & 500-1010  & 1.58 & 0.025 & 739 ($\times$3)   &  2,217     \\
 G140L (FUV) & 2020-12-29 12:37  & 1150-1730 (1425)  & 960-1440  & 0.60 & 0.025 & 887 ($\times$3)   &  2,661     \\
\enddata
\tablecomments{The observations were centered on PDS 70 using target coordinates (J2000)
   R.A. = 14$^{h}$ 08$^{m}$ 10.154$^{s}$, Decl. = $-$41$^{\circ}$ 23$'$ 52.58$''$.
   All exposures were obtained in ACCUM mode with the 52$''$ x 0.2$''$ slit.
   Target acquisition was performed using a 0.4 s STIS CCD image taken through
   the F28X50LP long-pass filter. Grating properties are from the STIS Instrument Handbook.
}
\tablenotetext{a}{Central wavelength in parentheses.}
\tablenotetext{b}{Resolving Power = $\lambda$/(2$\Delta$$\lambda$).}
\end{deluxetable}

\clearpage

\section{Observations and Data Reduction}

The {\em HST} STIS long-slit observations of PDS 70 were obtained
during two consecutive orbits on 29 December 2020 using  the Multi-Anode Micro-channel 
Array (MAMA) detectors. NUV and FUV spectra were obtained using the low-resolution 
G230L and G140L gratings respectively. Three exposures of equal length were 
obtained for each grating. Table 2 summarizes the observations and basic 
instrument properties.
Data were analyzed using PyRaf v. 2.1.6 and IRAF v. 2.7. 
Calibrated one-dimensional spectral files (x1d.fits) for the individual 
exposures provided by the STIS pipeline were combined into a single spectrum for 
each grating using tools in the Space Telescope Data Analysis System (STSDAS). Spectral
fitting and line property measurements were undertaken using STSDAS tools.
Fluxes were dereddened using the extinction law of Whittet et al. (2004)
assuming A$_{\rm V}$ = 0.05 mag for PDS 70.

\section{Results}

\subsection{Acquisition Image}

Target acquisition was performed using a STIS CCD image with a 0.4 s exposure
taken through the long-pass F28X50LP filter. This filter spans a wavelength
range of $\approx$5400 - 10,000 \AA ~with maximum sensitivity over
6000 - 7400  \AA. The raw acquisition image near PDS 70 is shown in Figure 1. 
In addition to 
PDS 70 a second faint source is visible at an offset of $\approx$2.5$''$  along 
P.A. $\approx$ 14$^{\circ}$. This source is found in the {\em Gaia} EDR3 catalog at 
magnitude $G$ = 15.504, about 4 mag fainter  than PDS 70 ($G$ = 11.606).
The {\em Gaia} EDR3 parallax of the faint source is 0.042$\pm$0.038 mas,
nearly consistent with zero. Its small parallax and apparent faintness
suggest it is a distant background object.
This faint source was not captured in the STIS grating slit 
(aperture P.A. = $-$134.35$^{\circ}$) and did not affect the UV spectra.

\begin{figure}[h]
\figurenum{1}
\includegraphics*[width=7.0cm,angle=0]{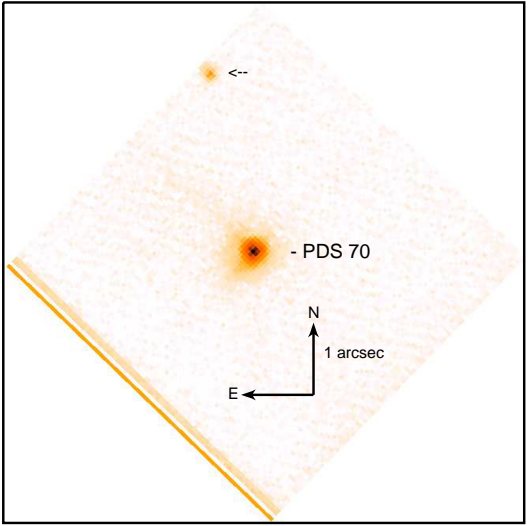} \\
\caption{Raw STIS CCD acquisition image of PDS 70 taken
through the F28X50LP filter showing a second faint source
({\em Gaia} EDR3 source\_id 6110141563309613184) located $\approx$2.5$''$ NE of PDS 70.
}
\end{figure}

\normalsize

\clearpage

\subsection{UV Spectra}
~The STIS FUV and NUV spectra obtained by combining the individual exposures
are shown in Figure 2.  Prominent emission lines are marked and
measured line fluxes are summarized in Tables 3 and 4 along with selected
NUV continuum flux density measurements taken over narrow wavelength
intervals where no obvious lines are visible or listed in the
Chianti Atomic Database\footnote{https://www.chiantidatabase.org}.

\subsubsection{Continuum}
Since PDS 70 is a late-type (K7) pre-main
sequence star, UV continuum emission is anticipated from the stellar
atmosphere. This will include  photospheric emission that
peaks in the visible range but extends downward into the NUV
plus continuum from hotter chromospheric and transition
region plasma in the upper atmosphere.
A question of interest is whether excess continuum emission is 
present in addition to that of the star's atmosphere.
Excess UV continuum is often seen in cTTS and is usually
attributed to an accretion shock (Johns-Krull et al. 2000; Ingleby
et al. 2013). In addition, collision and excitation of H$_{2}$
by energetic electrons can give rise to FUV continuum emission
via H$_{2}$ dissociation (Bergin et al. 2004; France, Yang, \& Linsky 2011).
However, even in accreting cTTS any NUV continuum emission originating
in an accretion shock is not easily distinguished from that of hot
plasma in the star's  upper atmosphere as discussed
by Kravtsova \& Lamzin (2003).

Figure 3 shows {\em dereddened} continuum flux density measurements of
PDS 70 in narrow wavelength intervals where no emission lines
are visible or listed in the Chianti Atomic Database.
Dereddening has very little effect assuming A$_{V}$ = 0.05 for PDS 70.
The photospheric continuum is clearly detected in the NUV
and a power-law fit extrapolates to zero photospheric flux density
at $\approx$2545 \AA~. This cutoff is consistent with predictions of
stellar photosphere models for K7 spectral type, e.g. the Pickles
K7V (T$_{\rm eff}$ = 3999 K) template in the STIS Exposure Time Calculator (ETC). 
Taking the ratio of
the dereddened photospheric continuum flux densities at
different wavelengths (Table 3) yields an estimate of the equivalent
NUV-derived blackbody temperature T$_{\rm eff(nuv)}$.
Specifically,  F$_{2922}$/F$_{3077}$ gives T$_{\rm eff(nuv)}$ = 4007 K
and F$_{2875}$/F$_{3077}$ gives a slighly higher value T$_{\rm eff(nuv)}$ = 4048 K.
These values are within 1\% - 2\%  of the value T$_{\rm eff}$ = 3972$\pm$37 K
obtained by Pecaut \& Mamajek (2016). This agreement is reasonably good
considering that STIS flux density measurements at the above wavelengths
are uncertain by 3\% - 5\% and A$_{V}$ is uncertain by a
factor of two (Table 1).

Significant NUV continuum is also present in the 1795 - 2561 \AA~ range
below the photospheric cutoff. But at the high end of this range
($\approx$2330 - 2550 \AA~) no reliable continuum measurements were obtained
due to many closely-spaced emission lines that cannot be individually
identified at the resolution of G230L, especially Fe II transitions.
Faint  emission that {\em may} be weak continuum is also measured in
the FUV over the range 1282 - 1515 \AA~. However, any real FUV continuum
emission that may be present is of low significance. Figure 4 
compares the gross, net, and background count rates over a narrow
FUV wavelength interval (1453 - 1454.9 \AA) where no emission 
lines are present. The ratio of gross (source$+$background) to background rates is
2.2/1.2 = 1.8 which is too low to qualify as a significant continuum detection.
But we do note here that near-constant FUV continuum emission
has been reported in some accreting cTTS (Kravtsova \& Lamzin 2003;
France et al. 2014).

In summary, the NUV continuum emission of PDS 70
longward of $\approx$2600 \AA~ is dominated by the
star's phototosphere. Below this cutoff the NUV continuum
is dominated by hotter thermal plasma
from the star's chromosphere and transition region. Any
excess FUV continuum emission that might be present 
is too faint to be clearly distinguished from MAMA background

\subsubsection{Emission Lines}
Measured line centroids agree with lab
wavelengths to within the absolute STIS wavelength accuracies
but many of the detected lines are closely-spaced blends that are
not clearly separated by the low-resolution STIS gratings. Maximum line
power temperatures (Table 3) lie within the range
log T$_{max}$ = 4.5 - 5.2 (K) as typical for chromospheric
and transition region plasma.  The lines with the highest T$_{max}$
are N V and Ne IV. The feature identified as Ne IV (1601.45 \AA) is quite 
faint and thus classified as a possible detection.
The brightest lines are the blended C IV doublet in G140L and 
the Mg II doublet in G230L. Density-sensitive carbon lines
are discussed further below (Sec. 3.4).

The H I Ly$\alpha$ line (1215.7 \AA) is
also brightly detected but is adversely affected by Earth's geocoronal
emission and interstellar absorption (resonant scattering of
photons out of the line-of-sight). 
Reconstructing the intrinsic Ly$\alpha$ line and its flux from the 
observed line requires {\em HST} medium resolution or echelle grating 
spectra capable of resolving the line profile
(e.g. Wood et al. 2005; Youngblood et al. 2022).
The low-resolution G140L spectrum analyzed here is insufficient for 
this purpose.

We compared the C IV doublet lines of the three exposures (Fig. 5) to determine
if any significant flux variability occurred. Only small differences 
were found with at most a
$\pm$6\% variation in the integrated C IV doublet flux for
the individual exposures compared to their mean value. The
measured (absorbed) C IV line flux in the combined spectrum
and range for the three exposures was
F$_{CIV}$ (1548$+$1550) = 1.33 (1.29 - 1.44) $\times$ 10$^{-14}$ 
ergs cm$^{-2}$ s$^{-1}$.

\underline{H$_{2}$ Lines}: 
Somewhat surprisingly, a search of the G140L spectrum at several
wavelengths where fluorescent H$_{2}$ lines are often detected in
accreting cTTS revealed five faint H$_{2}$ emission 
lines  as identified in Figure 6 and Table 4. 
Other known  H$_{2}$ lines whose lab wavelengths lie in the wings
of bright lines such as C IV may also be present but cannot be 
distinguished at G140L spectral resolution.
The presence of H$_{2}$ lines in the UV spectra of cTTS
is thought to be due mainly to excitation of H$_{2}$ near the star
radiatively pumped by stellar Ly$\alpha$ photons and subsequent 
fluorescent deexcitation (Ardila et al. 2002). 
Although most cTTS observed with {\em HST}
FUV gratings show fluorescent H$_{2}$ lines their detection 
in wTTS is rare. All five H$_{2}$ lines in the PDS 70 spectrum
have the same vibrational quantum number ($\nu'$ = 1)
for the excited electronic state and all but one have the same
rotational quantum number ($J'$ = 4) for the excited state.
The dereddened line fluxes are low compared to typical values
for cTTS (e.g. Fig. 5 of France et al. 2012 for [1,4] transitions).

The detection of H$_{2}$ lines provides evidence for molecular
gas in the vicinity of the star and STIS provides a moderate
constraint on its spatial extent.
As a  check for extended H$_{2}$ emission we extracted 
G140L spectra centered at offsets as close as $\pm$12 pixels
($\pm$0$''$.30) from the stellar trace using the MAMA default 
extraction width of 11 pixels ($\pm$0$''$.14) in the
spatial direction. At the distance of PDS 70 an 
offset of 0$''$.30$\pm$0$''$.14 corresponds to a projected 
separation of 34$\pm$16 au. 
No significant H$_{2}$ emission was seen in these off-source spectra 
at the wavelengths where on-source H$_{2}$ lines are visible. Thus
the H$_{2}$ lines originate within a projected separation of 
$\leq$34 au from the star.
This region contains the planets, part of the dust-depleted gap,
the region at $<$10 au where an inner disk may be present
(Long et al. 2018; Keppler et al. 2019), and of course the star itself.

To place tighter constraints on the location of the H$_{2}$
emission in PDS 70 a more sensitive FUV observation using 
higher spatial resolution imaging or higher spectral resolution
gratings capable of accurately measuring 
line centroids and resolving line shapes and widths is needed. 
If the H$_{2}$ emission originates in the disk close to the star 
then line centroid shifts should be negligible and excess line 
broadening from gas orbiting in the disk is expected. For H$_{2}$ emission 
originating in a molecular outflow significant line centroid
shifts relative to the stellar radial velocity are anticipated.
Stellar H$_{2}$ lines (e.g. from cool spots) would be unshifted
with broadening consistent with $v$sin$i$.

In {\em HST} COS and STIS FUV grating observations of a sample of
disked cTTS by France et al. (2012),  most stars revealed H$_{2}$ lines
broadened to FWHM $\approx$ 20 - 90 km s$^{-1}$ with little or no
velocity shifts. They concluded that the H$_{2}$
emission of these stars arises in the disk at small
separations of $\ltsimeq$ 3 au. Similarly, some cTTS in the 
sample observed with STIS by Herczeg et al. (2006) 
showed H$_{2}$ lines centered on the star's radial velocity.
But a few cTTS in the France et al. and Herczeg et al.
samples and of those observed by Ardila et al. (2002) 
revealed H$_{2}$ lines blueshifted up to a few
tens of km s$^{-1}$ indicative of low-velocity
molecular outflows. Blueshifted H$_{2}$ FUV lines
were also detected in STIS observations of the jet-driving 
cTTS RY Tau by  Skinner et al. (2018). Wide-angle H$_{2}$ FUV emission 
that may arise in a disk wind was also reported
for the jet-driving cTTS DG Tau by Schneider et al. (2013a,b).
Taken together, the above  results suggest that 
different regions contribute to fluorescent H$_{2}$ 
emission in cTTS and models that incorporate both 
disks and molecular outflows are required to 
explain the observations.

\begin{deluxetable}{lccccc}
\tabletypesize{\scriptsize}
\tablecaption{PDS 70 UV Emission Lines}
\tablehead{
           \colhead{Grating}                                   &
           \colhead{Name}                                       &
           \colhead{$\lambda_{lab}$\tablenotemark{\scriptsize a}}          &
           \colhead{$\lambda_{obs}$\tablenotemark{\scriptsize b}}                      &
           \colhead{log T$_{\rm max}$\tablenotemark{\scriptsize a}}                         &
           \colhead{Flux\tablenotemark{\scriptsize b}}  \\                     
           \colhead{}                         &
           \colhead{}                         &
           \colhead{(\AA~)}                   &
           \colhead{(\AA~)}                   &
           \colhead{(K)}                      &
           \colhead{(10$^{-15}$erg cm$^{-2}$ s$^{-1}$)}     
}
\startdata
G140L  & C III   & 1176.37          & 1176.5  & 4.8  & 2.42  \\
G140L  & Si III  & 1206.50          & 1206.6  & 4.7  & 5.39   \\
G140L  & C III   & 1247.38          & ...     & 4.9  & $\leq$0.08  \\
G140L  & N V     & 1238.82, 1242.81   & 1238.6    & 5.2  & 1.62\tablenotemark{\scriptsize c}  \\
G140L  & Si II, O I\tablenotemark{\scriptsize d}  & 1304.37,1304.86  & 1304.7    & 4.5  & 4.17   \\
G140L  & C II    & 1334.53,1335.71  & 1335.2    & 4.5  & 4.73    \\
G140L  & Si IV   & 1393.76          & 1393.5    & 4.9  & 4.57   \\
G140L  & Si IV   & 1402.77          & 1402.4    & 4.9  & 4.16   \\
G140L  & C IV    & 1548.19$+$1550.78  & 1548.7  & 5.0  & 13.3\tablenotemark{\scriptsize c} \\
G140L  & Fe II   & 1560.25, 1563.79   & 1561.8  & 4.5  & 2.4  \\
G140L  & Ne IV ?   & 1601.45            & 1601.49 & 5.2  & 0.9  \\
G140L  & He II   & 1640.34,1640.54    & 1640.2   & 4.9  & 5.83  \\
G140L  & C I     & 1656.93,1657.008\tablenotemark{\scriptsize e}   & 1656.9    & ...  & 2.64  \\
       &                 &           &            &      &      \\
G230L  & Si II   & 1816.93, 1817.45  & 1817.2 & 4.5   & 5.79  \\
G230L  & Si III  & 1892.03           & 1893.0 & 4.6   & 7.94   \\
G230L  & C III   & 1908.73            & 1909.1 & 4.8  & 5.00  \\
G230L  & C II   & 2326.12, 2327.65            & 2327.9       & 4.5  & 7.93     \\
G230L  & Fe II   & 2629.09, 2632.11          & 2629.3 & 4.5  & 20.5      \\
G230L  & Al II   & 2669.95        & 2670.1  & 4.5  & 3.10   \\
G230L  & Mg IIk        & 2796.35               & 2797.2 & 4.5  & 104.0    \\ 
G230L  & Mg IIh        & 2803.53               & 2804.0 & 4.5  & 70.8   \\
       &                 &           &            &      &      \\
G230L  & continuum     & ... & 2077.7$\pm$2.4   & ...  & 0.109 (0.125)\tablenotemark{\scriptsize f}     \\
G230L  & continuum     & ... & 2875.3$\pm$4.2   & ...  & 0.960 (1.047)\tablenotemark{\scriptsize f,g}     \\
G230L  & continuum     & ... & 2921.9$\pm$2.4   & ...  & 1.074 (1.170)\tablenotemark{\scriptsize f,g}     \\
G230L  & continuum     & ... & 2970.0$\pm$2.5   & ...  & 1.221 (1.328)\tablenotemark{\scriptsize f,g}     \\
G230L  & continuum     & ... & 3077.5$\pm$2.5   & ...  & 1.550 (1.682)\tablenotemark{\scriptsize f,g}
\enddata
\tablenotetext{a}{Lab wavelengths and maximum line power temperatures are from 
the Chianti atomic database (www.chiantidatabase.org) unless otherwise noted.}
\tablenotetext{b}{Measured wavelengths and observed line fluxes are based on Gaussian
fits and line fluxes are continuum-subtracted.  STIS MAMA absolute wavelength
accuracy is 0.5 - 1.0 pixel, or 0.3 - 0.6 \AA~ (G140L) and 0.8 - 1.6 \AA~(G230L).
For continuum measurements the wavelength uncertainty specifies the range over which 
the mean flux density was measured.}
\tablenotetext{c}{Sum of the two components based on a two-component Gaussian fit.}
\tablenotetext{d}{O I is geocoronal.}
\tablenotetext{e}{Lab wavelengths from NIST (physics.nist.gov/PhysRefData/Handbook/Tables/carbontable2.htm).}
\tablenotetext{f}{Mean of the observed flux density (units: 10$^{-15}$ erg cm$^{-2}$ s$^{-1}$ \AA$^{-1}$)
                 over the specified wavelength interval followed in
                 parentheses by dereddened values computed using the extinction law of Whittet et al. (2004)
                 and A$_{V}$ = 0.05.}
\tablenotetext{g}{Flux density is dominated by photospheric continuum.}
\end{deluxetable}

\clearpage

\begin{deluxetable}{lcccccc}
\tabletypesize{\scriptsize}
\tablecaption{PDS 70 H$_{2}$ FUV Emission Lines}
\tablehead{
           \colhead{Grating}                                   &
           \colhead{Transition\tablenotemark{\scriptsize a}}                                       &
           \colhead{[$\nu'$,$J'$]}            &
           \colhead{$\lambda_{lab}$}          &
           \colhead{$\lambda_{obs}$}          &
           \colhead{$\lambda_{pump}$}         &
           \colhead{Flux\tablenotemark{\scriptsize b}}  \\                     
           \colhead{}                         &
           \colhead{}                         &
           \colhead{}                         & 
          \colhead{(\AA~)}                    &
           \colhead{(\AA~)}                   &
           \colhead{(\AA~)}                   &
           \colhead{(10$^{-15}$erg cm$^{-2}$ s$^{-1}$)}     
}
\startdata
G140L  &  R(3) 1-6 & [1,4] & 1431.01   & 1431.1    & 1216.07  & 0.86 (0.97) \\
G140L  &  R(3) 1-7 & [1,4] & 1489.57   & 1489.6    & 1216.07  & 1.45 (1.64) \\
G140L  &  P(5) 1-6 & [1,4] & 1446.12   & 1445.8    & 1216.07  & 1.02 (1.15) \\
G140L  &  P(5) 1-7 & [1,4] & 1504.76   & 1504.6    & 1216.07  & 1.30 (1.46) \\
G140L  &  P(8) 1-6 & [1,7] & 1467.08   & 1467.3    & 1215.73  & 0.53 (0.60) \\
\enddata
\tablenotetext{a}{H$_{2}$ transitions are in the Lyman band system (Abgrall et al. 1993).
Vibrational and rotational quantum numbers in the upper electronic state are
[$\nu'$,$J'$] and in the lower electronic state [$\nu''$,$J''$]. The letter P denotes
$\Delta$$J$ = $-$1 and R denotes $\Delta$$J$ = $+$1 where 
$\Delta$$J$ = $J'$ $-$ $J''$. The number in parentheses is $J''$.
The two hyphenated numbers are the vibrational quantum numbers
$\nu'$ $-$ $\nu''$ in the upper and lower electronic states.}
\tablenotetext{b}{Observed fluxes are followed in
parentheses by dereddened values computed using the extinction law of Whittet et al. (2004)
and A$_{V}$ = 0.05}
\end{deluxetable}

\clearpage

\begin{figure}
\figurenum{2}
\includegraphics*[width=9.0cm,angle=-90]{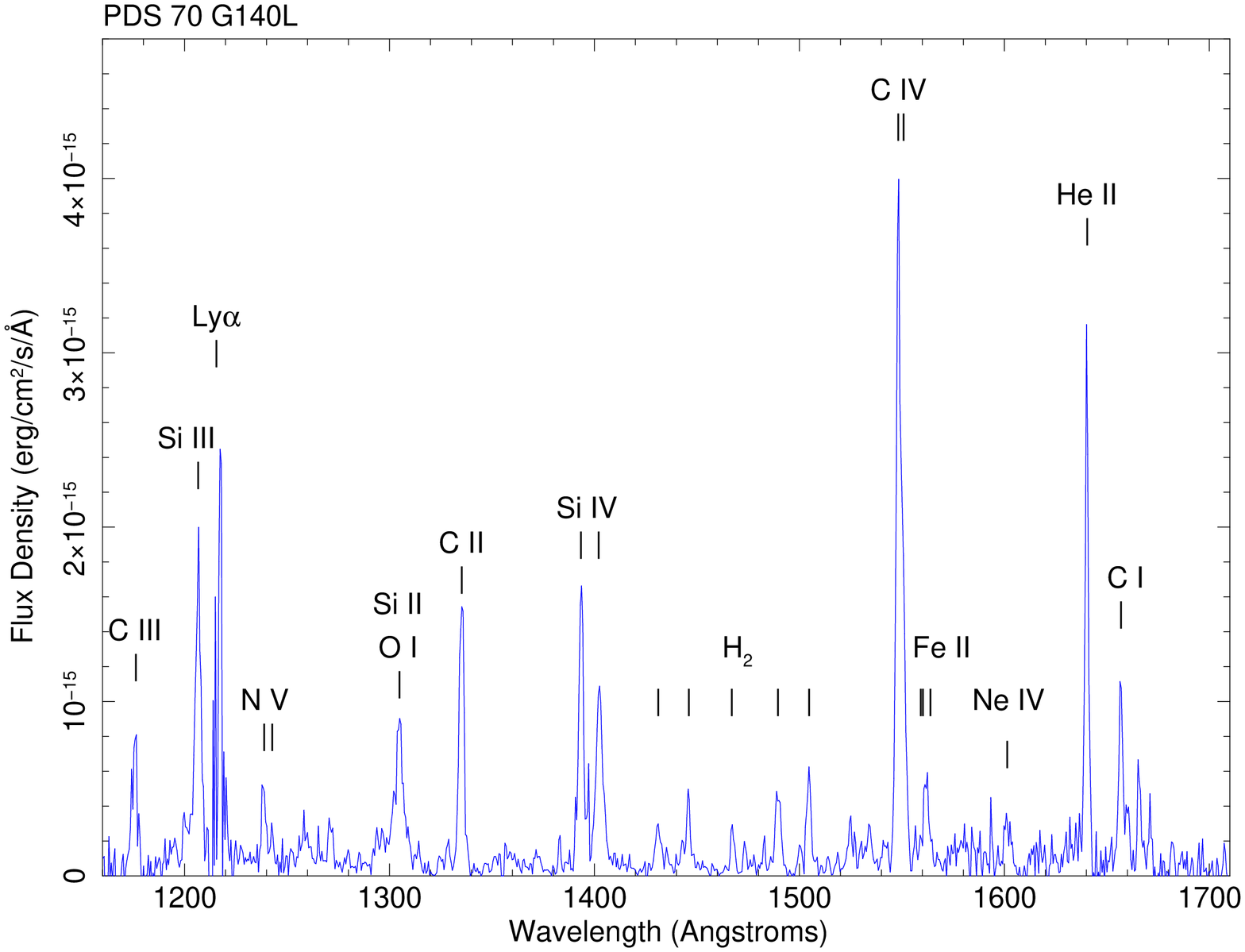} \\
\includegraphics*[width=9.0cm,angle=-90]{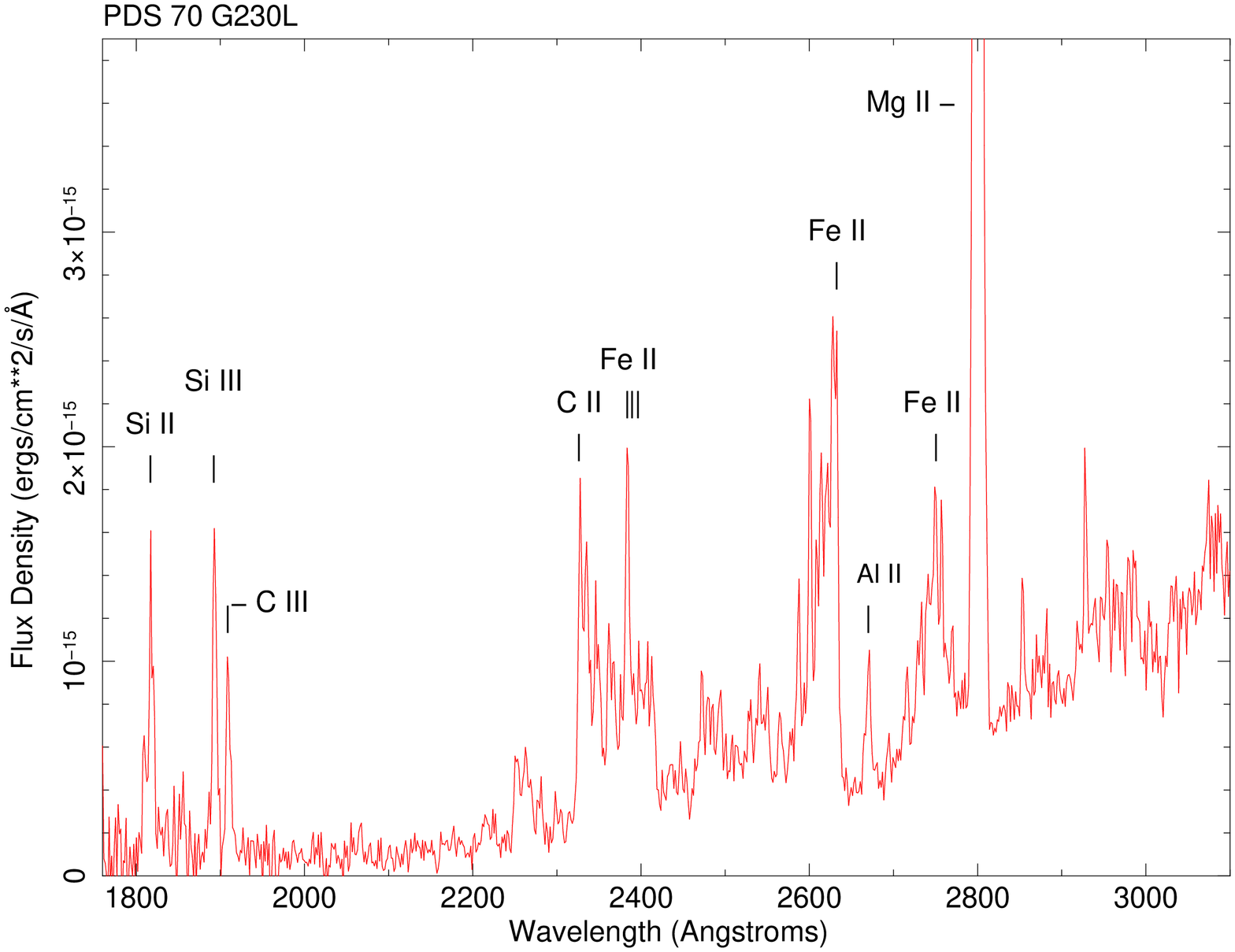}
\caption{STIS/MAMA G140L and G230L spectra of PDS 70 from the combined exposures
with prominent emission lines identified. The O I line in G140L is geocoronal.
}
\end{figure}

\begin{figure}
\figurenum{3}
\includegraphics*[width=7.0cm,angle=-90]{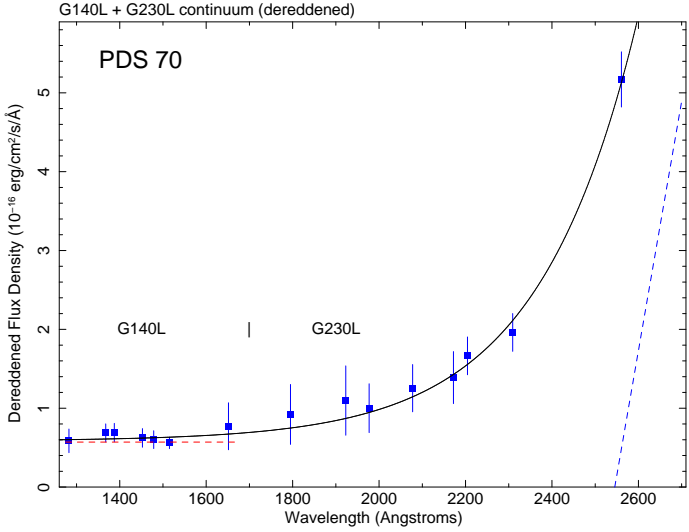} \\
\caption{PDS 70 G140L and G230L continuum flux densities measured in narrow wavelength
intervals and dereddened using A$_{V}$ = 0.05 and the extinction law of Whittet et al. (2004)
Flux densities are in units of 10$^{-16}$ erg/cm$^2$/s/\AA~.
The sloping blue line at far right is a power-law fit to four dereddened 
photospheric continuum measurements (Table 3) 
F$_{\lambda,dered}$ = $-$101.5 $+$ 0.1947$\lambda^{+0.7978}$ extrapolated to
zero flux density at $\lambda$ = 2545 \AA~.
The red dashed line at left shows the mean FUV MAMA background level (no dereddening)
at F$_{\lambda,bgd}$ = 5.42e-17 erg/cm$^{2}$/s/\AA~.
The curved line is a fit of 15 continuum data points (1282 - 2561 \AA~) using an 
exponential model with a constant offset of the form 
F$_{\lambda,dered}$ = $C$ $+$ $A$exp[($\lambda$ $-$ $\lambda_{0}$)/EW.
Best fit values are $C$ = 0.5836, $A$ = 0.2908, EW = 230.7,
$\lambda_{0}$ = 1926. The data and model become indistinguishable from background 
at $\lambda$ $<$ 1600 \AA~.
}
\end{figure}

\begin{figure}
\figurenum{4}
\includegraphics*[width=7.0cm,angle=-90]{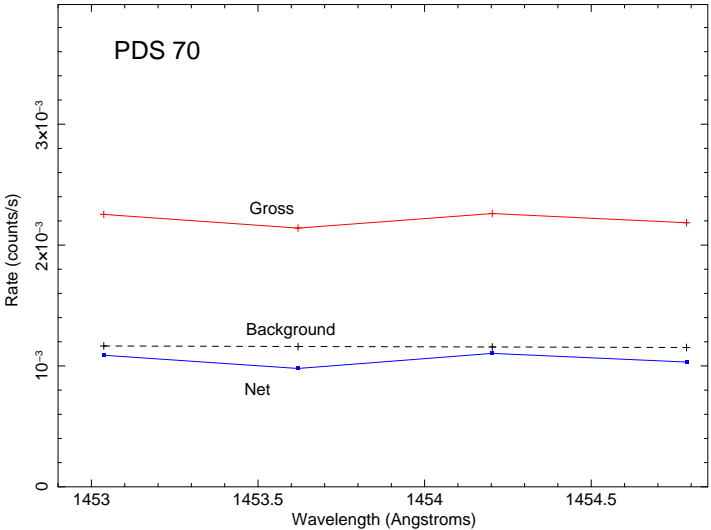} \\
\caption{PDS 70 G140L gross (= source$+$background), net (= source$-$background), and background 
count rates from the combined exposures in a narrow wavelength interval where no 
significant line emission  is expected. The background rate is based on the average of 
two background spectra extracted at offsets of $\pm$300 pixels ($\pm$7$''$.5) from the stellar trace.
Source and background rates were measured in spectra extracted with a width of 11 pixels in 
the cross-dispersion (spatial) direction.
}
\end{figure}

\begin{figure}
\figurenum{5}
\includegraphics*[width=9.0cm,angle=-90]{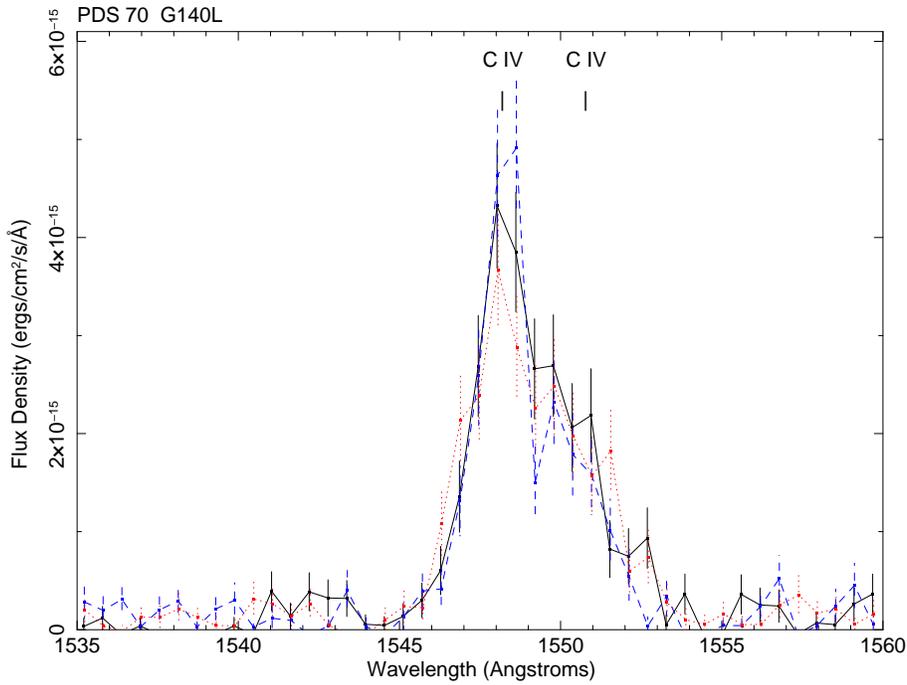} \\
\caption{
STIS/MAMA G140L spectra of PDS 70 for the three exposures showing the C IV doublet 
($\lambda_{lab}$ = 1548.19, 1550.78 \AA). Solid black: exposure 1, 
dashed blue: exposure 2, dotted red: exposure 3. 
Error bars are shown only for the first exposure.
}
\end{figure}

\begin{figure}
\figurenum{6}
\includegraphics*[width=9.0cm,angle=-90]{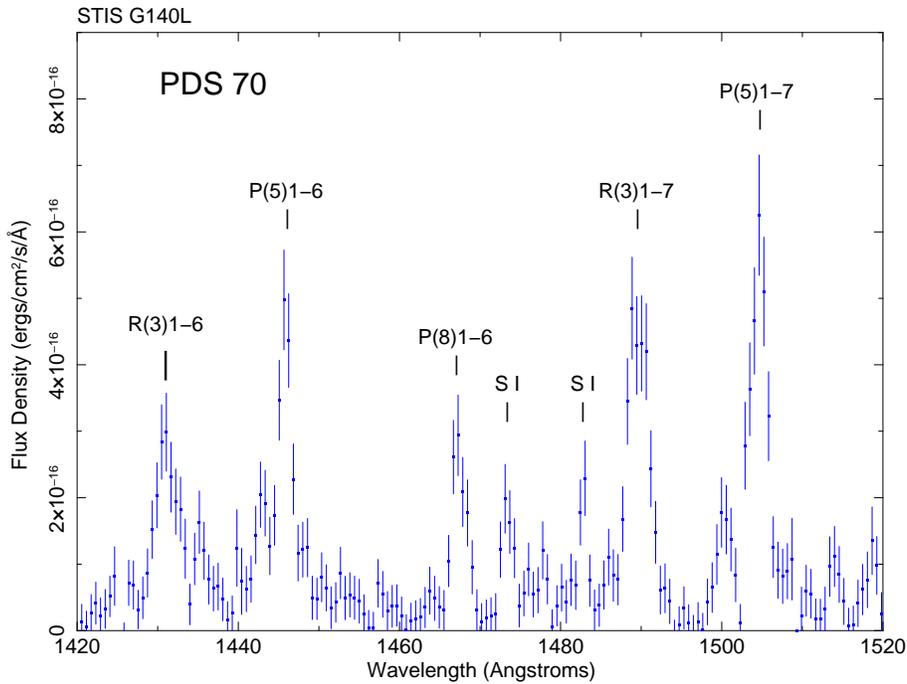} \\
\caption{STIS/MAMA G140L spectrum of PDS 70 from combined exposures
showing five prominent fluorescent H$_{2}$ lines. Tick marks show lab wavelengths of
the H$_{2}$ transitions (Table 4). Weak neutral sulfur lines 
are marked at 1474 and 1483 \AA.}
\end{figure}

\clearpage

\subsection{UV Luminosity}
\underline{FUV and NUV}:~
The FUV luminosity is dominated by line emission. To compute
L$_{FUV}$ we exclude Ly$\alpha$ (1216 \AA) and O I (1304.9$+$1306.0 \AA)
since Ly$\alpha$ is heavily absorbed by the interstellar medium and
both lines are contaminated by geocoronal
emission\footnote{https://www.stsci.edu/itt/APT\_help/STIS\_Cycle22/c06\_exptime6.html}.
Dereddening the other FUV line fluxes in Table 3 and summing gives
L$_{FUV,line}$(1170 - 1700 \AA) = 8.54 $\times$ 10$^{28}$ erg s$^{-1}$.
The largest contributor is the C IV doublet with a dereddened flux
F$_{C IV,dered}$(1548$+$1551) = 1.50 $\times$ 10$^{-14}$ erg cm$^{-2}$ s$^{-1}$
and L$_{C IV}$  = 2.26 $\times$ 10$^{28}$ erg s$^{-1}$.
If the near-constant exponential tail of the FUV continuum flux density
fit below 1700 \AA~ shown in Figure 3 is included but treated as an upper limit
to the FUV continuum then 
L$_{FUV,line+cont}$(1170 - 1700 \AA) $\leq$ 1.43 $\times$ 10$^{29}$ erg s$^{-1}$.
The NUV luminosity is dominated by the continuum, especially
above 2600 \AA~ where the stellar photosphere begins to contribute
significantly. Thus, the value obtained for L$_{NUV}$ depends
sensitively on the wavelength adopted for the upper limit
in the calculation. Using an upper limit of 3000 \AA~ gives
L$_{NUV,line+cont}$(1700 - 3000 \AA) = 1.2 $\times$ 10$^{30}$ erg s$^{-1}$.
The Mg $\rm{II}$ doublet contribution is
L$_{Mg II}$(2797$+$2804) = 2.88 $\times$ 10$^{29}$ erg s$^{-1}$.

\underline{EUV}:~Direct measurements of the EUV flux and luminosity
(0.013 - 0.1 keV) are only possible for the Sun and a
few nearby stars due to insterstellar absorption.
But  methods for estimating $L_{EUV}$ in late-type stars have been
proposed using $L_{X}$ as a reference.
For the solar-like star EK Dra (age $\sim$70 Myr) a ratio
$L_{EUV}$/$L_{X}$ $\approx$ 1 was found by Ribas et al. (2005).
Analysis of a large sample of mature late-type stars of ages
$\gtsimeq$100 Myr by Sanz-Forcada et al. (2011) provided a relation between
$L_{EUV}$ and $L_{X}$ but with large uncertainties in the fit coefficients.
Applying their fit to PDS 70 gives $L_{EUV}$/$L_{X}$ $\approx$ 4.
A similar ratio is obtained using the theoretical relations between
$L_{EUV}$, $L_{X}$, and age developed by Tu et al. (2015).
These estimates are subject to the caveat that the relations of 
Sanz-Forcada et al. and Tu et al. are not well-constrained observationally 
for young stars with ages $\ltsimeq$30 Myr such as T Tauri stars. 
Scaling relations of $L_{X}$ with age or rotation rate break down in 
young rapidly-rotating stars when coronal X-ray emission saturates at 
high levels of magnetic activity. X-ray saturation occurs at
$L_{X}$ $\approx$ 10$^{-3}L_{*}$ (G\"{u}del, Guinan, \& Skinner 1997).

{\em Swift} X-ray observations of PDS 70 analyzed by
Joyce et al. (2020) obtained an intrinsic (unabsorbed) X-ray flux
F$_{X,int}$(0.3-10 keV) = 3.4 $\times$ 10$^{-13}$ erg cm$^{-2}$ s$^{-1}$
or log L$_{X}$ = 29.71 erg s$^{-1}$ at a distance of 112.39 pc.
More extensive {\em XMM-Newton} observations in mid-2020
show that the X-ray luminosity of PDS 70 is variable and increases
by a factor of $\sim$3 or more during periods of high activity.
During low activity states (e.g. ObsIds 0863800201 and 0863800401),
our fits of archived {\em XMM-Newton} EPIC PN spectra give
F$_{X,int}$(0.2-7.5 keV) $\approx$ 4.18 $\times$ 10$^{-13}$ erg cm$^{-2}$ s$^{-1}$,
or log L$_{X}$ = 29.80 erg s$^{-1}$.
Taking this latter value as representative For PDS 70 we obtain 
$L_{X}$/$L_{*}$ = 10$^{-3.3}$ placing it close to the
expected saturation boundary.

\subsection{Density Sensitive Line Ratios}
Flux ratios of density sensitive lines can be used to estimate
the electron density n$_{e}$ in regions where the lines are formed.
The derived densities are sensitive to several factors including
abundances. However if ratios of different lines of the same element
are used the abundance uncertainty is avoided.
Ratios of C III transitions have been used to estimate n$_{e}$
in the solar atmosphere (Cook \& Nicolas 1979, hereafter CN79) 
and in some cTTS (Johns-Krull et al. 2000). This line forms at 
a maximum line power temperature log T$_{max}$ = 4.8 (K) which is 
within the range of the solar transition region.

The FUV spectrum of PDS 70 shows C III lines at 1176.4 \AA~ 
and 1908.7 \AA~ but the C III line at 1247.4 \AA~ is undetected (Table 3).
Dereddening the 1176.4 and 1908.7 \AA~ lines gives a
ratio F$_{dered}$(1176)/F$_{dered}$(1908) = 0.50.
Assuming these lines form at or near T$_{max}$ gives
log n$_{e}$ = 8.5 (cm$^{-3}$) using the results of 
CN79 evaluated on their log T = 4.75 curve.
However, they note that the 1176/1908  line ratio underestimates
n$_{e}$ by about a factor of five relative to other line ratios.
Applying a factor of 5 correction to the above gives
log n$_{e}$ = 9.2 (cm$^{-3}$). Again using the 
CN79 results the upper limit 
F$_{dered}$(1247)/F$_{dered}$(1908) $\leq$ 0.016
gives log n$_{e}$ $\leq$ 9.9 (cm$^{-3}$). 
Densities in the above range 
9.2 $\leq$ log n$_{e}$ $<$ 9.9 (cm$^{-3}$)
are typical of transition region plasma (Doschek 1997) and
provide confidence that the C III lines arise in 
the stellar atmosphere.

\clearpage

\section{Discussion}

\subsection{PDS 70 Accretion Rate and Luminosity}
Weak-lined T Tauri stars typically lack the accretion
signatures commonly seen cTTS
such as UV continuum excesses, broadened
high-temperature FUV lines (e.g. C IV), and fluorescent 
H$_{2}$ emission. As a representative case, Figure 7 shows 
that fluorescent H$_{2}$ lines are absent in the UV spectrum
of the wTTS HBC 427.
Nevertheless, some TTS do show accretion signatures such as
fluorescent H$_{2}$ emission even though other indicators are absent.
These include RECX-11 which is thought to be accreting at a low rate
$\dot{M}_{acc}$ $\leq$ 3 $\times$ 10$^{-10}$ M$_{\odot}$ yr$^{-1}$
(Ingleby et al. 2011; France et al. 2012). 

\begin{figure}
\figurenum{7}
\includegraphics*[width=9.0cm,angle=-90]{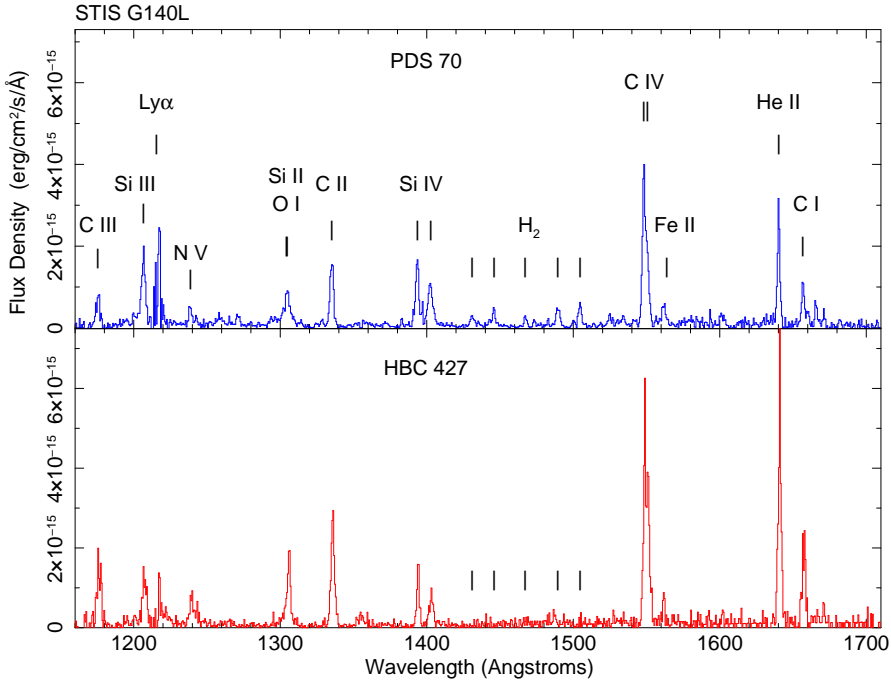} \\
\caption{STIS/MAMA G140L spectra of PDS 70 and the K7e-type wTTS
HBC 427 (= V397 Aur). The flux density
of HBC 427 has been scaled up by a factor
of 1.748 to correct for its greater distance (148.6 pc) than  PDS 70 (112.39 pc).
Error bars are not shown to facilitate comparison.
The HBC 427 G140L archive spectrum (O8NS07020) is from a 1425 s observation on
20 Jan. 2004 using the 52$''$$\times$0$''$.2 slit.
Note the lack of H$_{2}$ lines in the HBC 427  spectrum.
}
\end{figure}

The detection of FUV H$_{2}$ lines in PDS 70 supports previous 
claims that it is weakly accreting. To estimate the mass 
accretion rate  we use the relation 
log $\dot{M}_{acc}$ ($M_{\odot}$ yr$^{-1}$) = 
($-$5.4$\pm$0.2) $+$ (0.8$\pm$0.1) log(L$_{CIV}$/L$_{\odot}$)
given by Ardila et al. (2013). As a cautionary note, this relation 
was derived using a sample consisting mostly of cTTS with
considerable scatter in $\dot{M}_{acc}$
and is not well-calibrated at lower L$_{CIV}$ values typical of wTTS.
We  assume A$_{V}$ = 0.05 (0.025 - 0.10) for PDS 70, allowing for 
a factor of two uncertainty. Dereddening the observed flux using
this extinction gives 
F$_{CIV,dered}$(1548$+$1551) = 1.5 (1.41 - 1.69) $\times$ 10$^{-14}$
ergs cm$^{-2}$ s$^{-1}$ and
log(L$_{CIV}$/L$_{\odot}$) = $-$5.23 ($-$5.25 - $-$5.18).
The inferred accretion rate is then 
$\dot{M}_{acc}$ = 2.6 (0.5 - 14.8) $\times$ 10$^{-10}$ $M_{\odot}$ yr$^{-1}$
where the range in parentheses accounts for the uncertainties in
A$_{V}$ and the regression relation coefficients.
The above range is  consistent with the previous estimates of
$\dot{M}_{acc}$ obtained by Manara et al. (2019) and 
Thanathibodee et al. (2020) given in Section 1.

Based on an analysis of FUV spectra of a sample of 91 pre-main sequence stars,
Yang et al. (2012) obtained the following best-fit relation between
L$_{CIV}$ and accretion luminosity:
log(L$_{CIV}$/L$_{\odot}$) = $-$2.766 $+$ 0.877 log(L$_{acc}$/L$_{\odot}$).
Using the same range of L$_{CIV}$ values for PDS 70 as above that take
A$_{V}$ uncertainty into account, this
relation gives log(L$_{acc}$/L$_{\odot}$) = $-$2.81 ($-$2.83 - $-$2.75)
or L$_{acc}$ = 5.93 (5.66 - 6.80) $\times$ 10$^{30}$ erg s$^{-1}$.
 
As a consistency check on the above we invoke the relation
L$_{acc}$ = $\xi$G$\dot{M}_{acc}$M$_{*}$/R$_{*}$
where $\xi$ = [1 - (R$_{*}$/$r_{i}$)] and 
$r_{i}$ is the radius from which the accreting gas 
originates (Hartmann et al. 2016).
Using the PDS 70 stellar parameters from Table 1 this becomes
\begin{equation}
{\rm L}_{acc} = \xi(7.3 \times 10^{30})\left[\frac{\dot{M}_{acc}}{10^{-10}~ {\rm M_{\odot}~ yr^{-1}}}\right]  ~ {\rm erg~ s^{-1}}.
\end{equation}

An additional constraint is that $r_{i}$ $\leq$ $r_{co}$ = 6.4 R$_{*}$
or equivalently $\xi$ $\leq$ 0.844 where $r_{co}$ is the disk corotation 
radius for an assumed stellar rotation period P$_{rot}$ = 3.03 d (Thanathibodee et al. 2020). 
Combining this constraint with the above constraints on $\dot{M}_{acc}$ and L$_{acc}$ leads 
to the more restricted range
0.9 $\leq$ [$\dot{M}_{acc}$/10$^{-10}$ M$_{\odot}$ yr$^{-1}$] $\leq$ 1.1 and
3.4 $\leq$ $r_{i}$/R$_{*}$ $\leq$ 6.4
(i.e. 0.705 $\leq$ $\xi$ $\leq$ 0.844).
For comparison, the magnetospheric
model of Thanathibodee et al. (2020) obtained a disk truncation 
radius $r_{i}$/R$_{*}$ = 4.0 which lies within the above range.

\subsection{Planet Irradiation by the Star}
The disk gas model based on the ALMA observations of
Keppler et al. (2019) leads to very high column 
densities at the distance of the inner planet,
effectively blocking stellar X-ray and UV radiation. 
The disk gas mass density in the midplane at a 
distance $r$ from the star is
$\rho(r)$ = 0.4 $\Sigma_{gas}(r)$/$H(r)$ 
where $\Sigma_{gas}(r)$ is the gas surface density
and $H(r)$ the disk height radial profile
(eqs. [6] and [7] of Keppler et al. 2019).

Renormalizing the Keppler et al. (2019)  results to
a fiducial distance of $r$ = 1 au
gives $\Sigma_{gas}(r)$ = $\Sigma_{0}$$r_{au}^{-1}$
and $H(r)$ = $H_{0}$$r_{au}^{+1.25}$ where
$\Sigma_{0}$ $\approx$ 10$^{2}$ g cm$^{-2}$,
$H_{0}$ = 0.04 au, and $r_{au}$ is in units of au.
We have neglected the 
exponential decay term in the expression for $\Sigma_{gas}(r)$ 
used by Keppler et al.  since it is near unity
for small values of $r$ in the inner disk where
stellar X-ray and UV radiation are heavily absorbed.
The adopted radial profile $\Sigma_{gas}(r)$ $\propto$ $r^{-1}$
is only valid out to $r$ $\approx$ 10-11 au, beyond which
the gas density falls off much more rapidly
in the gap (Fig. 9 of Keppler et al. 2019).
The behavior of $\Sigma_{gas}(r)$ at $r$ $\ltsimeq$ 5 au
is not well-constrained observationally since this
innermost region is not spatially resolved by ALMA.
We thus leave $\Sigma_{0}$ as a free parameter below.

The disk gas number density is $n_{H}(r)$ = $\rho(r)$/($\mu$m$_{p}$)
where m$_{p}$ is the proton mass and we assume a mean
molecular weight $\mu$ = 2.3 (amu) for cold disk gas dominated by H and
He in a ratio by number of H:He = 10 and H predominantly
in molecular form as H$_{2}$.  Integrating $n_{H}(r)$ 
through the disk from inner radius $r_{i}$ to outer
radius $r_{f}$ gives the equivalent
neutral H column density 
N$_{\rm H}$ = $\int_{r_{i}}^{r_{f}}$$n_{\rm H}(r)$d$r$.
For X-ray photons of characteristic energy $E_{X}$ = kT$_{X}$ the
optical depth is
$\tau_{X}$ = $\sigma$($E_{X}$)N$_{\rm H}$ where
$\sigma$($E_{X}$) = $\sigma_{0}$($E_{X}$/1 keV)$^{-p}$ cm$^{2}$
is the X-ray photoelectric absorption cross section. We
adopt $\sigma_{0}$ = 2.27 $\times$ 10$^{-22}$ cm$^{2}$ and $p$ = 2.485
(Igea \& Glassgold 1999; Shang et al. 2002).
For the X-ray photon energy  $E_{X}$ $\approx$ 1 keV typical of PDS 70
in low-activity states the above gives
unit optical depth $\tau_{X}$ = 1 at
N$_{\rm H}$ = 4.4 $\times$ 10$^{21}$ cm$^{-2}$.

Formally, the lower limit $r_{i}$ on the integration to 
determine N$_{\rm H}$ should be taken as the inner truncation 
radius of the gas disk. This value is not well-known but
the estimates given above yield values of a few stellar
radii, or much less than 1 au.
Since $\Sigma_{gas}(r)$ is not well-determined
close to the star we set $r_{in}$ = 5 au 
for the integration. Then integrating out to $r_{f}$ = 10 au 
provides a {\em lower limit} on the absorption of X-rays by 
{\em pre-gap} disk gas.

Integrating with the above limits  gives a column density 
N$_{\rm H}$($r$ $\leq$ 10 au)  
$\geq$ 1.6 $\times$ 10$^{23}$$\Sigma_{0}$ cm$^{-2}$.
In order for X-rays to penetrate this inner disk region ($\tau_{X}$ $<$ 1)
a surface density $\Sigma_{0}$ $\ltsimeq$ 10$^{-2}$ g cm$^{-2}$
at $r$ = 1 au is required. This is at least four orders of magnitude 
less than implied by the inner (unperturbed) disk gas model 
of Keppler et al. (2019).  Thus, based on current knowledge of 
the disk gas profile as constrained by ALMA, the inner planet at $r$ = 22 au 
is well-shielded from keV X-rays. The same conclusion holds for 
stellar UV radiation, which is more heavily absorbed than X-rays 
by virtue of its lower energy range and 
larger photoelectric absorption cross section (Bruderer et al. 2009).

\subsection{Disk Photoevaporative Mass Loss and Dissipation}
XUV irradiation of disk gas leads to mass loss via a photoevaporative wind.
This process, along with accretion onto the star (and planets, in the case
of PDS 70) eventually depletes the disk gas and accretion ceases. 
X-ray, EUV, and FUV radiation can influence photoevaporative mass loss
rates ($\dot{M}_{pe}$) to varying degrees and detailed models must consider 
their relative importance. Theoretical estimates of $\dot{M}_{pe}$ are sensitive 
to the adopted stellar and disk model parameters, the evolutionary state
of the disk (i.e. early primordial disks versus evolved gapped disks),
and to what degree relevant physical processes are included. Many 
physical processes can affect the disk temperature, density, and chemical 
distributions as reviewed by Woitke (2015).
A key quantity entering into mass loss rate calculations is
the disk gas surface density profile $\Sigma_{gas}(r)$ which changes as
the disk evolves (Fig. 7 of Gorti, Dullemond, \& Hollenbach 2009). 
For most T Tauri stars including PDS 70 the 
behavior of $\Sigma_{gas}(r)$ at small radii $r$ $\ltsimeq$ 5 au is not
well-constrained observationally due to limits on telescope 
angular resolution.

More than one study suggests that EUV radiation is less effective
in driving gaseous disk evolution than X-ray or the combination of
X-ray$+$FUV radiation (Gorti et al. 2009; Owen et al. 2012). Furthermore, 
the effects of FUV radiation are more important at early evolutionary
stages when accretion rates are high and L$_{FUV}$ is enhanced by
accretion shocks (Gorti et al. 2009). Since PDS 70 is now only weakly 
accreting it is likely that X-ray radiation is the dominant factor driving its 
photoevaporative mass loss, although that was not necessarily
the case at earlier stages.

Most models predict that the X-ray driven mass loss rate
$\dot{M}_{pe,x}$ scales nearly linearly with L$_{X}$ and  
softer X-ray spectra lead to higher values of $\dot{M}_{pe,x}$, 
all other factors being equal (Gorti et al. 2009; Ercolano et al. 2009). 
This point is relevant to PDS 70 whose  X-ray spectrum in
low-activity states is quite soft for a wTTS, peaking
near 1 keV and then falling off rapidly with 
negligible flux above 2 keV. Some accreting cTTS
such as TW Hya also have similar soft X-ray spectra
(Kastner et al. 1999; 2002). 

To obtain a rough estimate of the disk mass loss rate for PDS 70
we use the results of Alexander et al. (2006). Their formulation
is appropriate for later stages of disk evolution when the 
inner disk is sufficiently cleared for direct stellar radiation
to drive disk mass loss. This may be the case for the highly
evolved PDS 70 disk but as noted above its disk gas surface density 
profile close to the star ($r$ $\ltsimeq$ 5 au) is not yet 
well-constrained observationally. 

Assuming the inner disk is sufficiently cleared to be directly
affected by the star's ionizing radiation the photoevaporative 
mass loss rate is (eq. [3] of Alexander et al. 2006)
\begin{equation}
\dot{M}_{pe}(r < R_{out}) = 9.69 \times 10^{-10} \mu C \left[\frac{H/R}{0.05}\right]^{-1/2}\left[\frac{R_{in}}{3~{\rm au}}\right]^{1/2} \left[\frac{\Phi}{10^{41}~ {\rm s}^{-1}}\right]^{1/2}~~M_{\odot} yr^{-1}.
\end{equation} 
In the above we  have used fiducial values given by Alexander et al.  to compute the 
leading constant. The values of $R_{in}$, $R_{out}$ are the inner and outer disk radii,
$\mu$ is the mean mass per particle (amu), $C$ = [1 - ($R_{in}$/$R_{out}$)$^{0.42}$],
$H/R$ is the ratio of disk scale height to radius, 
and $\Phi$ is the ionizing
photon flux. As noted by Alexander et al., most of the mass loss
occurs at radii near the gravitational radius $r_{g}$ beyond which gas is no
longer gravitationally bound to the star. Using $r_{g}$ = GM$_{*}$/$c_{s}^2$
where $c_{s}$ $\approx$ 10 km s$^{-1}$ is the sound speed assuming 
the escaping gas temperature is $T_{gas}$ $\sim$ 10$^{4}$ K (Owen et al. 2012) 
gives $r_{g}$ $\approx$ 7 au for PDS 70. However, in hydrodynamic models the 
gas begins to escape at smaller radii and the effective value of
$r_{g}$ is reduced (e.g. Font et al. 2004). Thus, most of the mass loss
for PDS 70 is expected to occur inside 10 au. This is likely the case
since the PDS 70 disk gas surface density falls off 
rapidly for $r$ $>$ 10 au in the gap. We thus take  
$R_{out}$ = 10 au for the mass loss calculation.
The value of $R_{in}$ is not well-constrained
but values in the range $R_{in}$ = 1 - 3 au yield similar mass loss rates
and we adopt $R_{in}$ = 1 au.
For PDS 70, the disk models of Keppler et al. (2019)
give (H/R) = 0.05 at 3 au and (H/R) = 0.07 at 10 au so we just use the fiducial
value (H/R) = 0.05 in the above equation.

For the ionizing photon flux $\Phi$ we include both EUV and X-ray emission,
assuming L$_{EUV}$ $\approx$ L$_{X}$ = 6 $\times$ 10$^{29}$ erg s$^{-1}$.
Adopting mean photon energies E$_{EUV}$ = 0.1 keV and E$_{X}$ = 1 keV
gives $\Phi_{EUV}$ = 3.75 $\times$ 10$^{39}$ s$^{-1}$ and
$\Phi_{X}$ = 3.75 $\times$ 10$^{38}$ s$^{-1}$. However, each
primary X-ray photon spawns multiple secondary ionizations
which effectively increase $\Phi_{X}$. If E$_{ion}$ $\approx$ 0.037 keV
is the energy required to create an ion pair (G\"{u}del 2015) then
E$_{X}$/E$_{ion}$ $\approx$ 27 and the adjusted ionization rate
is $\Phi_{X}$ = 1.0 $\times$ 10$^{40}$ s$^{-1}$.
We thus obtain $\Phi$ = $\Phi_{EUV}$ $+$ $\Phi_{X}$ =
1.375 $\times$ 10$^{40}$ s$^{-1}$. Inserting this into the above equation
gives $\dot{M}_{pe}$($r <$ 10 au) = 1.3 $\times$ 10$^{-10}$$\mu$ M$_{\odot}$ yr$^{-1}$
where $\mu$ $\approx$ 1.35 for atomic gas and  $\mu$ $\approx$ 2.3 for molecular gas.
This mass loss rate is comparable to the PDS 70 mass accretion rate (Sec. 4.1).
The value of $\dot{M}_{pe}$ is not very sensitive to the assumed value of 
L$_{EUV}$ which, as noted previously, is not observationally determined.
If L$_{EUV}$ is increased by a factor of 4 then $\dot{M}_{pe}$ only 
increases by a factor of 1.35.

For comparison with the above, the studies of Ercolano et al. (2009) and 
Drake et al. (2009) focus on X-ray driven mass loss
and assume a disk that is in hydrostatic equilibrium and 
stellar parameters that are a reasonably close match for 
PDS 70. Their adopted X-ray spectrum includes soft
emission E$_{X}$ $<$ 1 keV and assumes a mean temperature log T$_{X}$ = 7.2 K 
(kT$_{X}$ = 1.37 keV), only slightly higher than for PDS 70 during 
low activity (kT$_{X}$ $\approx$ 1 keV).
The Drake et al. models consider a range of L$_{X}$ values that brackets
PDS 70. For log L$_{X}$ = 29.8 ergs s$^{-1}$ their results give 
$\dot{M}_{pe,x}$ $\approx$ 5 $\times$ 10$^{-10}$ M$_{\odot}$ yr$^{-1}$.
This is likely an overestimate for PDS 70 since their adopted disk model
assumes a surface density profile $\Sigma(r)$ with a smooth power-law
decline based on  D'Alessio et al. (1998), whereas the 
PDS 70 disk is heavily perturbed with a wide low surface density gap
centered near 22 au. Taking the likelihood of an overestimate into
account, the predicted mass loss rate of Drake et al.  may not differ much from 
that based on the formulation of Alexander et al. (2006) given above, that is
$\dot{M}_{pe}$ $\sim$ 10$^{-10}$ M$_{\odot}$ yr$^{-1}$.
But other studies such as the hydrodynamic mass loss
models of generic disks described  by Owen et al. (2012)
predict mass loss rates an order of magnitude larger
than above. Taken together, these results suggest
$\dot{M}_{pe}$ $\gtsimeq$ $\dot{M}_{acc}$ for PDS 70.
In this case disk mass loss competes with or even 
starves accretion as discussed by Drake et al. (2009).

\subsection{Disk Lifetimes}
The predicted disk lifetime from theoretical models depends on 
several factors including the assumed initial (primordial) disk mass
and viscosity. As such, disk lifetime predictions are heavily 
model-dependent. Other relevant factors are binarity and
environmental effects such as OB star winds and strong UV radiation 
fields in massive star-forming regions.
To briefly summarize previous work, the models of Alexander et al. (2006) 
predict $t_{disk}$ $\sim$ 6 - 8 Myr for an assumed ionizing flux
$\Phi$ = 10$^{42}$ s$^{-1}$, significantly higher than PDS 70
at current epoch. The study of Gorti et al. (2009) considered the combined 
effects of X-ray, EUV, and FUV radiation and obtained  
$t_{disk}$ $\sim$ 2 - 6 Myr. The lower value corresponds to
a soft X-ray spectrum with significant emission below 1 keV
and stronger photoevaporation than harder spectra.
Their predictions assumed  log L$_{X}$ = 30.3 erg s$^{-1}$, 
about 3 times larger than PDS 70 at current epoch. 
The empirical study of Bertout et al. (2007) based on derived 
ages of TTS in Taurus-Auriga obtained an average disk 
lifetime $t_{disk}$ = 4(M$_{*}$/M$_{\odot}$)$^{0.75}$ Myr.
Their model does take mass accretion into account but
not photoevaporation. 
A similar study of the Lupus association by Galli et al. (2015)
deduced an average lifetime 
$t_{disk}$ = 3(M$_{*}$/M$_{\odot}$)$^{0.55}$ Myr.
Average disk lifetimes for specific star-forming regions provide
a useful reference but significant variation can also be present. 
The study of Taurus-Auriga disks by 
Armitage et al. (2003) reveals a large dispersion in
disk lifetimes. Most disks have dissipated at stellar
ages of $\sim$6 Myr but some survive longer. The reason
for the large dispersion is not known but variations in
initial disk mass is one possible explanation.

The above results span a rather large range but suggest 
that most TTS disks do not survive longer than $\sim$6 Myr,
the estimated age of PDS 70 (Table 1). We thus conclude
that the PDS 70 disk is near the end of its expected lifetime 
and in the final stages of being cleared. PDS 70 thus 
serves as a rare  example of the terminal stage of TTS 
disk evolution and as the host star of a formative planetary system.

\clearpage

\section{Summary}
 {\em  HST} STIS FUV and NUV spectra of PDS 70 reveal emission lines 
formed at chromospheric and transition region temperatures. Stellar
continuum is detected but any weak FUV continuum excess that might be
present from accretion shocks is of low significance. The detection 
of several FUV fluorescent H$_{2}$ lines provides evidence for 
molecular gas in the vicinity of the star that is potentially 
feeding weak accretion as the inner disk drains. XUV
irradiation of the remaining disk gas is predicted to be driving a
weak photoevaporative disk wind whose mass loss rate is comparable 
to or greater than the inferred accretion rate 
$\dot{M}_{pe}$ $\gtsimeq$   $\dot{M}_{acc}$ $\sim$ 10$^{-10}$ M$_{\odot}$ yr$^{-1}$.
The combined effects of photovaporation and residual accretion onto 
the star and planets portend the approaching end of the disk's
lifetime. As the disk clears the planets will be more directly
exposed to the star's ionizing radiation.

\begin{acknowledgments}
This work was supported by {\em HST} award HST-GO-16290 issued by the
Space Telescope Science Institute (STScI) and is based on observations
made with the NASA/ESA Hubble Space Telescope, operated by
the Association of Universities for Research in Astronomy, Inc. under
contract with NASA. 
This work has utilized data in the {\em XMM-Newton} Science Archive
and data analysis products including STSDAS and PyRAF produced by the STScI,
and HEASOFT developed and maintained by HEASARC at NASA GSFC.
\end{acknowledgments}

\vspace{5mm}
\facilities{{\em Hubble Space Telescope (STIS)}}

\vspace{5mm}

\clearpage


\clearpage

\begin{thebibliography}{}
\bibitem[Abgrall et al. (1993)]{abg93}
         Abgrall, H., Roueff, E., Launay, F., Roncin, J.-Y., \& Subtil, J.-L.  1993,
          A\&AS, 101, 273
\bibitem[Alexander et al. (2006)]{ale06}
         Alexander, R.D., Clarke, C.J., \& Pringle, J.E. 2006,
         \mnras, 369, 229 
\bibitem[Ardila et al. (2002)]{ard02}
         Ardila, D.R., Basri, G., Walter, F.M., Valenti, J.A., \& Johns-Krull, C.M. 2002,
         \apj, 566, 1100
\bibitem[Ardila et al. (2013)]{ard13}
         Ardila, D.R., Herczeg, G.J., Gregory, S.G. et al. 2013,
         \apjs, 207, 1
\bibitem[Armitage et al. (2003)]{arm03}
         Armitage, P.J., Clarke, C.J., \& Palla, F. 2003,
         \mnras, 342, 1139
\bibitem[Bergin et al. (2004)]{ber04}
         Bergin, E., Calvet, N., Sitko, M.L. et al. 2004,
         \apj, 614, L133
\bibitem[Bertout et al. (2007)]{ber07}
         Bertout, C., Siess, L., \& Cabrit, S. 2007,
         \aap, 473, L21
\bibitem[Bruderer et al. (2009)]{bru09}
        Bruderer, S., Doty, S.D., \& Benz, A.O. 2009,
        \apjs, 183, 179
\bibitem[Christiaens et al. (2019)]{chr19}
         Christiaens, V., Cantalloube, F., Casassus, S., Price, D.J.,, Absil, O., 
         Pinte, C., Girard, J., \& Montesinos, M. 2019, \apjl, 877, L33
\bibitem[Cook \& Nicolas (1979)]{coo79}
         Cook, J.W. \& Nicolas, K.R. 1979, \apj, 229, 1163 (CN79)
\bibitem[D'Alessio et al. (1998)]{dal98}
         D'Alessio, P., Cant\'{o}, J., Calvet, N., \& Lizano, S. 1998,
         \apj, 500, 411
\bibitem[Doschek (1997)]{dos97}
         Doschek, G.A. 1997, \apj, 476, 903
\bibitem[Drake et al. (2009)]{dra09}
         Drake, J.J., Ercolano, B., Flaccomio, E., \& Micela, G.  2009,
         \apj, 699, L35 
\bibitem[Ercolano et al. (2009)]{erc09}
         Ercolano, B., Clarke, C.J., \& Drake, J.J. 2009,
         \apj, 699, 1639 
\bibitem[Font et al. (2004)]{fon04}
         Font, A.S., McCarthy, I.G., Johnstone, D., \& Ballantyne, D.R. 2004,
         \apj, 607, 890
\bibitem[France et al. (2014)]{fra14}
         France, K., Schindhelm, E., Bergin, E.A., Roueff, E., \& Abgrall, H. 2014,
         \apj, 784, 127
\bibitem[France et al. (2012)]{fra12}
         France, K., Schindhelm, E., Herczeg, G.J. et al. 2012,
         \apj, 756, 171
\bibitem[France et al. (2011)]{fra11}
         France, K., Yang, H., \& Linsky, J.L. 2011,
         \apj, 729, 7
\bibitem[Galli et al. (2021)]{gal21}
         Galli, P.A.B., Bertout, C., Teixeira, R., \& Ducourant, C. 2015,
         \aap, 580, 26
\bibitem[Gorti et al. (2009)]{gor09}
        Gorti, U., Dullemond, C.P., \& Hollenbach, D. 2009,
        \apj, 705, 1237 
\bibitem[Gregorio-Hetem \& Hetem (2002)]{gre02}
         Gregorio-Hetem, J. \& Hetem Jr., A. 2002, \mnras, 336, 197
\bibitem[G\"{u}del (2015)]{gud15}
         G\"{u}del, M. 2015, EpJ Web of Conferences, 102, 00015
\bibitem[G\"{u}del et al. (1997)]{gud97}
         G\"{u}del. M., Guinan, E.F., \& Skinner, S.L. 1997, 
         \apj, 483, 947
\bibitem[Haffert et al. (2019)]{haf19}
        Haffert, S.Y., Bohn, A.J., de Boer, J., Snellen, I.A.G., Brinchmann, J.,
        Girard, J.H., Keller, C.U., \& Bacon, R. 2019, Nature Ast., 3, 749
\bibitem[Hartmann et al. (2016)]{har16}
        Hartmann, L., Herczeg, G., \& Calvet, N. 2016,
        \araa, 54, 135
\bibitem[Herczeg et al. (2006)]{her06}
        Herczeg, G.J., Linsky, J.L., Walter, F.M., Gahm, G.F., \& Johns-Krull, C.M. 
        2006, \apjs, 165, 256
\bibitem[Igea \& Glassgold (1999)]{ige99}
         Igea, J. \& Glassgold, A.E. 1999, \apj, 518, 848
\bibitem[Ingleby et al. (2011)]{ing11}
        Ingleby, L., Calvet, N., Bergin, E. et al. 2011,
        \apj, 743, 105
\bibitem[Ingleby et al. (2013)]{ing13}
        Ingleby, L., Calvet, N., Herczeg, G.  et al. 2013,
        \apj, 767, 112
\bibitem[Isella et al. (2019)]{ise19}
        Isella, A., Benisty, M., Teague, R., Bae, J., Keppler, M., 
        Facchini, S., \& P\'{e}rez, L. 2019, \apjl, 879,L25
\bibitem[Johns-Krull et al. (2000)]{joh00}
         Johns-Krull, C.M., Valenti, J.A., \& Linsky, J.L. 2000, 
         \apj, 539, 815
\bibitem[Joyce et al. (2020)]{joy20}
        Joyce, S.R.G., Pye, J.P., Nichols, J.D., Page, K.L., Alexander, R.,
        G\"{u}del, M., \& Metodieva, Y. 2020, 
        \mnras, 491, L56
\bibitem[Kastner et al. (1999)]{kas99}
        Kastner, J.H., Huenemoerder, D.P., Schulz, N.S., \& Weintraub, D.A. 1999,
        \apj, 525, 837
\bibitem[Kastner et al. (2002)]{kas02}
        Kastner, J.H., Huenemoerder, D.P., Schulz, N.S., Canizares, C.R., 
        \& Weintraub, D.A. 2002,
        \apj, 567, 434
\bibitem[Keppler et al. (2018)]{kep18}
        Keppler, M., Benisty, M., M\"{u}ller, A. et al. 2018,
        \aap, 617, A44
\bibitem[Keppler et al. (2019)]{kep19}
        Keppler, M., Teague, R., Bae, J. et al. 2019, 
        \aap, 625, A118
\bibitem[Kiraga (2012)]{kir12}
        Kiraga, M. 2012, Acta Astronomica, 62, 67
\bibitem[Kravtsova \& Lamzin (2003)]{kra03}
         Kravtsova, A.S. \& Lamzin, S.A. 2003, Astron. Lett., 29, 612
\bibitem[Long et al. (2018)]{lon18}
         Long, Z.C., Akiyama, E., Sitko, M. et al. 2018, 
         \apj, 858, 112
\bibitem[Manara et al. (2019)]{man19}
         Manara, C.F., Mordasini, C., Testi, L., Williams, J.P., Miotello, A.,
         Lodato, G., \& Emsenhuber, A. 2019, \aap, 631, L2
\bibitem[M\"{u}ller et al. (2018)]{mul18}
         M\"{u}ller, A., Keppler, M., Henning, Th. et al. 2018,
         \aap, 617, L2
\bibitem[Owen et al. (2012)]{owe12}
         Owen, J.E., Clarke, C.J., \& Ercolano, B. 2012, 
         \mnras, 422, 1880 
\bibitem[Pecaut \& Mamajek (2016)]{pec16}
         Pecaut, M.J. \& Mamajek, E.E. 2016, \mnras, 461, 794
\bibitem[Ribas et al. (2005)]{rib05}
         Ribas, I., Guinan, E.F., G\"{u}del, M., \& Audard, M. 2005,
         \apj, 622, 680
\bibitem[Sanz-Forcada et al. (2011)]{sfo11}
         Sanz-Forcada, J., Micela, G., Ribas, I. et al.  2011,
         \aap, 532, A6
\bibitem[Schneider et al. (2013a)]{schn13a}
          Schneider, P.C., Eisl\"{o}ffel, J.,  G\"{u}del, M., G\"{u}nther, H.M.,
             Herczeg, G., Robrade, J., \& Schmitt, J.H.M.M. \ 2013a, \aap, 550, L1
\bibitem[Schneider et al. (2013b)]{schn13b}
          Schneider, P.C., Eisl\"{o}ffel, J.,  G\"{u}del, M., G\"{u}nther, H.M.,
             Herczeg, G., Robrade, J., \& Schmitt, J.H.M.M. \ 2013b, \aap, 557, A110
\bibitem[Shang et al. (2002)]{sha02}
         Shang, H., Glassgold, A.E., Shu, F.H., \& Lizano, S. 2002, \apj, 564, 853
\bibitem[Skinner et al. (2018)]{ski18}
         Skinner, S.L., Schneider, P.C., Audard, M., \& G\"{u}del, M. 2018,
         \apj, 855, 143
\bibitem[Thanathibodee et al. (2020)]{tha20}
        Thanathibodee, T., Molina, B., Calvet, N. et al. 2020,
        \apj, 892, 81
\bibitem[Tu et al. (2015)]{tu15}
         Tu, L., Johnstone, C.P., G\"{u}del, M., \& Lammer, H. 2015,
         \aap, 577, L3
\bibitem[Whittet et al. (2004)]{whi04}
         Whittet, D.C.B., Shenoy, S.S., Clayton, G.C., \& Gordon, K.D. 2004,
         \apj, 602, 291
\bibitem[Woitke (2015)]{woi15}
         Woitke, P. 2015, EPJ Web of Conferences, 102, 00011
\bibitem[Wood et al. (2005)]{woo05}
         Wood, B.E., Redfield, S., Linsky, J.L., M\"{u}ller, H.-R., \& Zank, G.P. 2005,
         \apjs, 159, 118
\bibitem[Yang et al. (2012)]{yan12}
         Yang, H., Herczeg, G.J., Linsky, J.L. et al. 2012,
         \apj, 744, 121
\bibitem[Youngblood et al. (2022)]{you22}
         Youngblood, A., Pineda, J.S., Ayres, T., France, K., Linsky, J.L., 
         Wood, B.E., Redfield, S., \& Schlieder, J.E. 2022, \apj, 926, 129
\bibitem[Zhou et al. (2021)]{zho21}
         Zhou, Y., Bowler, B.P., Wagner, K.R. et al. 2021, \apj, 161, 244  
\end{thebibliography}
\end{document}